# Design of multifunctional color routers with Kerker switching using generative adversarial networks

Jiahao Yan[1]*, Dayu Zhu[2], Yanjun Bao[1], Qin Chen[1], Baojun Li[1], & Wenshan Cai[2]*

1. *Guangdong Provincial Key Laboratory of Nanophotonic Manipulation, Institute of Nanophotonics, Jinan University, Guangzhou 511443, China*

2. *School of Electrical and Computer Engineering, Georgia Institute of Technology, Atlanta, Georgia 30332, USA*

*Corresponding author: jhyan@jnu.edu.cn & wcai@gatech.edu

**To achieve optoelectronic devices with high resolution and efficiency, there is a pressing need for optical structural units that possess an ultrasmall footprint yet exhibit strong controllability in both the frequency and spatial domains. For dielectric nanoparticles, the overlap of electric and magnetic dipole moments can scatter light completely forward or backward, which is called Kerker theory. This effect can expand to any multipoles and any directions, re-named as generalized Kerker effect, and realize controllable light manipulation at full space and full spectrum using well-designed dielectric structures. However, the complex situations of multipole couplings make it difficult to achieve structural design. Here, generative artificial intelligence (AI) is utilized to facilitate multi-objective-oriented structural design, wherein we leverage the concept of "combined spectra" that consider both spectra and direction ratios as labels. The proposed generative adversarial network (GAN) is named as DDGAN (double-discriminator GAN) which discriminates both images and spectral labels. Using trained networks, we achieve the simultaneous design for scattering color and directivities, RGB color routers, as well as narrowband light routers. Notably, all generated structures possess a footprint less than 600x600 nm indicating their potential applications in optoelectronic devices with ultrahigh resolution.**

**Keywords: color routers; Kerker theory; narrowband routing; generative adversarial network; inverse design**

# 1. INTRODUCTION

Nanoscale light manipulation has long been a focal point of research in the field of optics and photonics. However, when devices shrink to the nanoscale, the collective effects observed in metasurfaces and photonic crystals are no longer applicable. Instead, maximizing the optical control capabilities of individual nanostructures becomes crucial. This need is particularly urgent in applications such as displays, imaging, and spectral imaging, which require nanoresonators as pixels capable of ultra-high spatial resolution and efficient light manipulation.[1-3] For example, in the application of optical structure-assisted CMOS image sensors, single element nanostructures need to simultaneously act as color filters and color routers.[4-6] Unfortunately, constructing optical pixels with subwavelength footprints remains a formidable challenge. Although prior studies have reported color routers based on dielectric or plasmonic nanoantennas,[7-12] their focus has been primarily on interpreting phenomena rather than achieving precise design. More recently, submicron-sized color routers have been demonstrated through intricate structural designs.[13-14] However, the inherent structural complexity of these designs introduces fabrication challenges. Consequently, the quest to achieve the most extensive spectral and spatial optical functionalities with the highest spatial resolution has become a critical concern.

For a single dielectric nanostructure, optical resonances (such as Mie, anapole, BIC, etc.) based on the electromagnetic multipole coupling offer solutions for frequency domain tailoring.[15-18] Additionally, the coupling between electromagnetic multipoles through the generalized Kerker effect presents the possibility of tailoring light directionally within the spatial domain.[19-25] By harnessing the rich multipole couplings of dielectric nanostructures, it becomes feasible to simultaneously achieve desired spectral and directional tailoring. However, current research on the Kerker effect remains focused on specific mode couplings, leaving certain issues unresolved. For instance, challenges persist in achieving designability in both the frequency and spatial domains, realizing multi-channel routing, and enabling light routing within the narrowest wavelength ranges. The difficulty lies in the complexity of designing multi-objective-oriented mode couplings.

The design of optical structures (metasurfaces or individual nanoantennas) mostly starts from fixed structures based on prior experience, followed by time-consuming trial-and-error parameter sweeping. However, since a limited number of structural parameters fails to provide sufficient degrees of freedom, achieving multi-objective designs for color routers proves challenging. Thanks to the rapid development of machine learning and deep learning, some optical designs that require great wisdom and complex mechanisms can be realized using AI,[26-30] especially generative AI. Notably, image-based generative networks offer a high degree of design freedom and have played a significant role in optical design endeavors, such as unit cells in metasurfaces.[31-38]

In this study, we present a novel approach based on generative adversarial networks (GANs) for designing single-element multifunctional color routers. Given the complexity of coupling between electromagnetic multipole modes, the spectrum of a dielectric resonator represents a complex combination of Fano and Lorentz lineshapes. The significant difference between scattering spectra and one-hot classification labels renders the classic conditional GAN (c-GAN) inappropriate. Consequently, we propose the use of a double-discriminator GAN (DDGAN) where the discriminators evaluate not only the authenticity of images (structures) but also assess the label (spectral) loss. This strongly supervised model evaluates both spectra and structures during the training process. The DDGAN referred WGAN-GP (Wasserstein GAN + Gradient Penalty) and was inspired by the idea of ACGAN (Auxiliary Classifier GAN). In addition, through the one-time spectrum collection of multiple detectors at multiple wavelength ranges, we created "combined spectra" for training with high degree of freedom on slicing and combining, which is suitable for multi-objective predictions.

Using the trained networks, we successfully achieve inverse designs for color routers in three different scenarios. Firstly, we achieve the simultaneous design of up scattering, down/up (D/U), left/right (L/R), and front/back (F/B) direction ratios. This enables the creation of structural color displays with hidden multiplex information within narrowband directional switching channels. Secondly, we design multi-wavelength radiation patterns by considering full-space scattering, resulting in RGB color routers with an ultrasmall footprint. Lastly, we focus on designing narrowband routers to achieve the highest switching rate of lateral scattering ($\Delta L/R_{max}$) at desired wavelengths. Theoretically, we utilize spectral and spatial multipole decomposition to explain the amplitude, phase, and shape of each mode in directional switching channels. Our findings highlight the importance of a narrowband mode with the correct amplitude and radiation pattern, which we term "Kerker switching." Thanks to the proposed DDGAN approach, we identify an ideal Kerker switching structure with a high switching rate and a large switching range. Notably, we demonstrate the significance of the electrical octupole (EO) mode as a key component of Kerker switching, a factor that was overlooked in other works. These discoveries present new design concepts for complex light manipulation at the nanoscale and pave the way for miniaturized optoelectronic devices.

## 2. RESULTS AND DISCUSSION

To understand the Kerker switching, we first studied the backward and forward scattering of silicon (Si) nanodisks with fixed height of 200 nm and swept diameters from 100 nm to 600 nm. Classical Kerker theory studies the directivity between forward and backward scattering arising from the coupling between electric and magnetic dipoles. To study further the generalized Kerker effect, larger structures with multipole modes are preferred. For clarity, we donate forward and backward as D and U, respectively. In Fig. 1a, the up scattering spectra (Fig. S1a) and down scattering spectra (Fig. S1b) are combined as logarithmic direction ratios ($lg(Scat_D/Scat_U)$), and the direction ratios in the wavelength range from 400 to 900 nm are presented. The contour lines of direction ratios show a redshift proportional to the increase in diameter. This behavior is similar to that observed in up and down scattering. Notably, some switching regions can be observed where the direction ratio dramatically changes if diameter or wavelength slightly adjusts. To reveal the switching rate, we also plot the differential form ($\Delta lg(Scat_D/Scat_U)$) in Fig. 1b where three main switching channels (A, B, and C) can be observed. The Si nanodisk with a diameter of 430 nm was chosen to study the three channels simultaneously. In Fig. S1c and d, the differential up and down scattering spectra are provided to guarantee that these three channels also have significant variation of scattering intensities and are not from trivial fluctuations. The scattering spectra and the direction ratios are presented together in Fig. 1c for the Si nanodisk with a diameter of 430 nm, and the three switching channels are marked. Using the multipole decomposition of arbitrary structures,[39, 40] we calculated the scattering cross section contributed by electric dipole (ED), magnetic dipole (MD), electric quadrupole (EQ), magnetic quadrupole (MQ), EO, and magnetic octupole (MO). The key modes that drive the Kerker switching at A, B, and C channels turn out to be the combined effect of MQ and EO, MD alone, and EQ alone, respectively. The schematic diagram in Fig. 1d indicates the coordinate axes, the polarization direction, and the incident direction. The 3D radiation patterns at both ends of each switching band are also presented in Fig. 1d and Fig. S2 to give a visualized understanding on the rapid up-down switching. Since the multipole decomposition of full-space cross section is not enough to describe the directional behavior, we proposed spatial multipole decomposition after considering the radiated far-field of each mode, and the radiation patterns at XZ-plane and YZ-plane are presented in Fig. 1e and Fig. S3, respectively. Both amplitude and phase are important to understand the scattering directivity. Importantly, the scattering parities are opposite for multipoles with the same type and adjacent orders (e.g., ED and EQ) and for multipoles with the same order and different types (e.g., ED and MD).[25] Therefore, in Fig. 1e, we marked whether in-phase or out-of-phase of the dominated modes. At A channel, the switching on of MQ and EO mode at λ=626 nm cancels the EQ mode in upward direction; at B channel, the switching on of MD mode at λ=759 nm cancels the ED mode in upward direction; and at C channel, the switching on of EQ mode at λ=827 nm cancels the ED mode in upward direction.

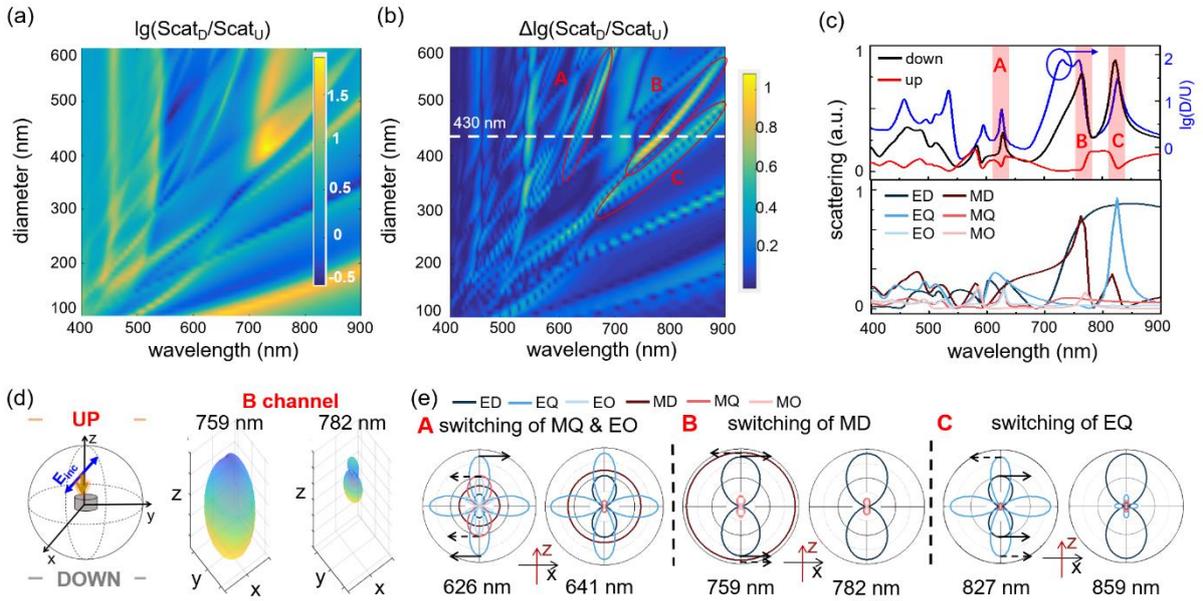

**Figure 1. The Kerker switching of axial scattering.** (a) Visualization of logarithmic direction ratios (lg(Scat$_D$/Scat$_U$)) of a Si nanodisk changing with wavelengths and diameters. (b) The differential of logarithmic direction ratios (Δlg(Scat$_D$/Scat$_U$)) varied with wavelengths and diameters. Red circles mark the A, B, and C switching channels, and the white dashed line indicates the chosen diameter (430 nm). (c) Up and down scattering and the corresponding lg(Scat$_D$/Scat$_U$) of the chosen Si nanodisk (d=430

nm), followed by the spectral decomposition of scattering cross section. (d) A schematic diagram showing the polarization direction, the XYZ axes, and the definition of Up and Down, and the 3D radiation patterns of B channel switching from 759 nm to 782 nm. (e) Radiation patterns contributed by each mode at three channels. The wavelengths and key switching modes are noted, and arrows are used to indicate the phase differences.

If we only consider axial scattering (up and down), the Kerker switching is easy to design by parameter sweeping and multipole analysis. However, the situation is more complex for lateral (left and right) scattering as shown in Fig. S4, where the diameter and height of Si nanodisk are fixed at 430 nm and 200 nm, respectively, and symmetry breaking is introduced through fan-shaped notch with varied degree. Similarly, we also presented the differential and logarithmic form of L/R direction ratios ($\Delta lg(Scat_L/Scat_R)$) (Fig. 2a) to find the switching channels. The contour of extreme values is irregular indicating the complex mode evolution. The schematic diagram in Fig. 2b gives the definition of angle $\varphi$, the L/R scattering, and the polarization direction. Another question of symmetry breaking nanodisks is the simultaneous degeneration of D/U directivities as shown in Fig. S5, indicating the difficulties on simultaneous design. In Fig. 2c, the angle $\varphi$ is fixed at 120º for describing the three switching channels (A, B, and C). For the symmetry-breaking nanodisk, the spatial and spectral distribution of each mode are strongly influenced. Therefore, there is no direct relevance between the direction ratio and the multipole decomposition of scattering cross section. To uncover the origin of lateral scattering switching, we calculated the radiation patterns of electromagnetic multipoles at the XY-plane, along with the corresponding phase information for left and right, as depicted in Fig. 2d and Fig. S6. For instance, at C channel, we observed a significant change in both the pattern and amplitude of the MD mode, indicating that the Kerker switching arises from a simultaneous dipole flip and dipole moment change. The complete radiated far-field under spherical coordinates ($E_\theta$, $E_\varphi$, and $E_r$) were calculated (Fig. S7). Furthermore, the phase information for $E_\varphi$ was obtained at $\varphi=0º$ (left) and 180º (right) and visualized in vector diagrams (Fig. 2d and Fig. S6). From the vector diagrams of C channel, we can conclude the activation of MD mode at 833 nm leads to in-phase coupling on the left and out-of-phase coupling on the right.

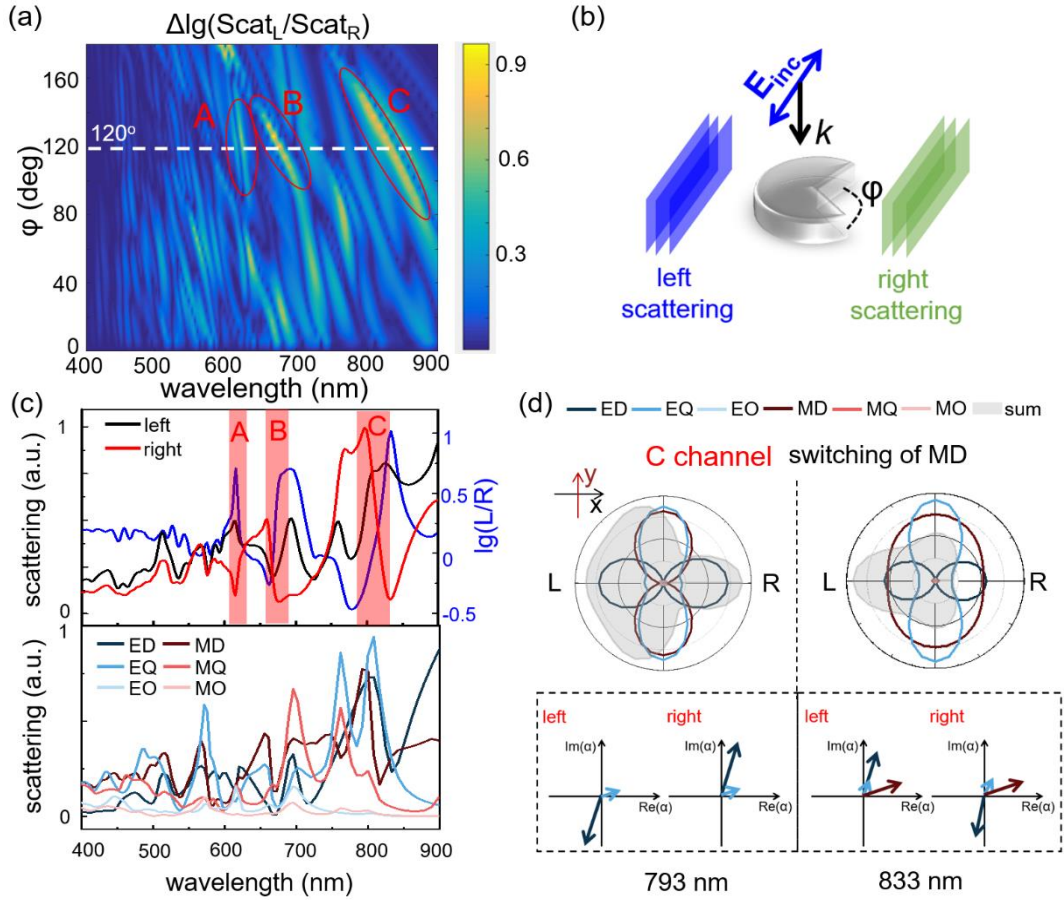

**Figure 2. The Kerker switching of lateral scattering.** (a) The differential logarithmic direction ratios changing with wavelengths and notch angles. The diameter and height of the Si nanodisk are fixed at 430 nm and 200 nm, respectively. Red circles mark the A, B, and C switching channels, and the white dashed line indicates the chosen notch angle (120°). (b) The definition of polarization direction, left (right) scattering, and the notch angle φ. (c) Left and right scattering, the logarithmic L/R ratio, and the mode-resolved scattering cross section of the chosen symmetry-breaking Si nanodisk (φ=120°). (d) XY-plane mode-resolved radiation patterns of C channel due to the switching of MD from 793 nm to 833 nm, and vector diagrams of left and right directions at each wavelength are provided.

Through the aforementioned analysis, it becomes evident that the electromagnetic multipoles of arbitrary Si structures are highly complex. Therefore, designing a structure with a specific optical response, both spatially and spectrally, poses a significant challenge when considering multipole coupling. Furthermore, conducting parameter-sweeping simulations without prior knowledge of multipole mode evolution can be time-consuming. To address these challenges, we employed the proposed DDGAN (Fig. 3) for inverse design, encompassing combination, training, and prediction processes. For the combination part (Fig. 3a), we first established training dataset consisting 30,000 structural designs (ellipses, rectangles, L-shapes, arcs, crosses, parallel dimers, and vertical dimers) as shown in Fig. S8. All structures are drawn in an $800 \times 800$ nm$^2$ region and converted to $64 \times 64$ matrix for training. The height of Si structures is 200 nm and no substrate is set which is consistent with the analysis in Fig. 1 and 2. Using finite-difference time-domain (FDTD) simulations, scattering spectra were collected over a wide wavelength range (400-1600 nm) in six directions (U, D, L, R, F, and B). Besides the scattering spectra, we calculated the direction ratios (D/U, L/R, and F/B) and the wavelength-dependent radiation angles ($\varphi_{max-xz}$, $\varphi_{max-yz}$) after considering all six planes. All these spectra and directional information form as a database, and combined spectra are formed through slicing the desired wavelength ranges and combining as vectors. It should be noticed that the specific data sieving needs to be done for every type of combined spectra to exclude useless data. For example, the direction ratio that is invariable across all wavelengths has no application value.

Through putting the selected images with corresponding spectral labels into the DDGAN as shown in Fig. 3b, the generator (*G*) and discriminator (*D*) networks can be trained. Our proposed DDGAN is based on WGAN-GP, where Wasserstein distance is used as value function to avoid possible vanishing gradients. Moreover, conventional WGAN needs careful tuning of the clipping threshold to deal with gradient vanishing or exploding. Therefore, in WGAN-GP, a soft version of the constraint with a penalty on the gradient norm for random samples $\hat{x} \sim P_{\hat{x}}$ is introduced. The discriminator loss is as follow:

$$L = \mathop{E}_{\tilde{x} \sim P_g}\left[D(\tilde{x})\right] - \mathop{E}_{x \sim P_r}\left[D(x)\right] + \lambda \mathop{E}_{\hat{x} \sim P_{\hat{x}}}\left[\left(\left\|\nabla_{\hat{x}} D(\hat{x})\right\|_2 - 1\right)^2\right] \qquad (1)$$

where $P_{\hat{x}}$ is the sampling uniformly along straight lines between pairs of points sampled from the data distribution $P_r$ and the generator distribution $P_g$. $\lambda = 10$ is the gradient penalty factor. Compared with WGAN-GP, modifications in DDGAN are as follow: 1. the dot product of combined spectrum and latent vector rather than the concatenation is fed into the Generator. 2. A label loss layer is added in the Discriminator and plus the mean absolute error (MAE) in Eq. 1 with a label loss factor ($\mu$).

The optimized DDGAN architectures through hyperparameter tuning are shown in Fig. S9. The training process includes two steps: first, the "fake" images are generated through the input of combined spectra (S) multiplied with a latent vector (Z). The latent vector is formed by normalized random numbers, which enables the one-to-many structural design. Second, both generated images with S and real images with S are fed into the Discriminator, where the value function $\mathop{E}\limits_{x \sim P_r}[D(x)] - \mathop{E}\limits_{\tilde{x} \sim P_g}[D(\tilde{x})]$ is maximized to enhance its discernibility. At the same time, the label (spectral) loss $\mu \frac{\sum_n |S - D'(\tilde{x})|}{n}$ is minimized to guarantee the strong supervision. Here, $D'$ means the other form of the Discriminator's output with the same dimension of labels (S). The optimization of Generator is achieved via maximizing $\mathop{E}\limits_{\tilde{x} \sim P_g}[D(\tilde{x})]$ in order to generate convincing images that deceive the discriminator. Consequently, the Generator and Discriminator with opposite objectives bring in competition and interactional improvement via the optimization of value functions.

Fig. 3c indicates the prediction and evaluation process: first, the target spectra (including test dataset and hand-drawn dataset) are combined and packed together as the input for trained Generator. Secondly, the generated structures are put into FDTD simulators to generate the same type of simulated combined spectra and then compare with the target combined spectra for accuracy evaluation. To confirm the advantages of our proposed DDGAN, the training processes were also conducted using DCGAN and WGAN-GP architectures with extensive hyperparameter tuning. The optimized parameters we used as well as the MAEs are concluded in Table S1. Notably, the combined spectra containing 160-point up scattering and 160-point D/U ratio were used, so the validation errors are presented separately as $MAE_{UP}$ and $MAE_{D/U}$. It turns out that DDGAN has the best performance with $MAE_{UP}$ and $MAE_{D/U}$ equal 0.118 and 0.052, respectively, while the $MAE_{UP}$ and $MAE_{D/U}$ equal 0.195 (0.171) and 0.105 (0.082) using DCGAN (WGAN-GP).

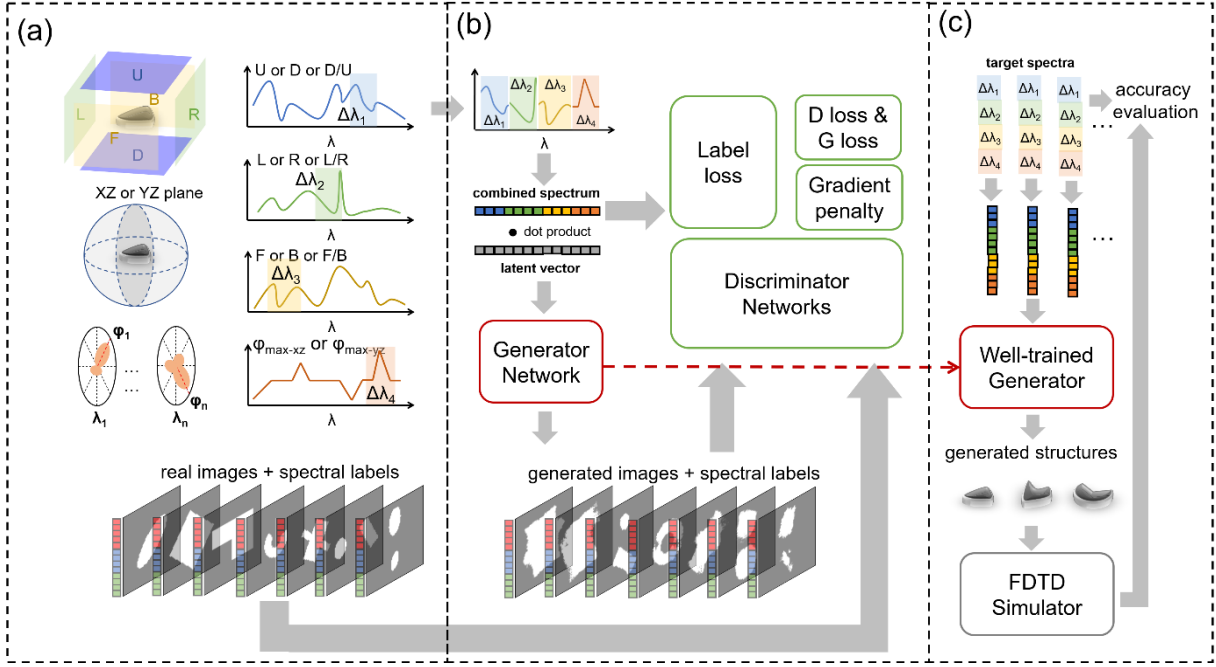

**Figure 3. The workflow and architecture of proposed GAN model.** (a) The generation of combined spectra dataset from spectral slicing and combination of U, D, L, R, F, B scattering spectra and D/U, L/R, F/B ratios. In addition, the radiation information calculated from six-plane scattering is also considered as spectral labels. (b) The training process of DDGAN through a generator and double discriminators. (c) The prediction and validation processes: 1. Input targeted combined spectra into well-trained generator; 2. Input generated structures into FDTD simulator; 3. Decompose targeted and simulated combined spectra, and evaluate accuracy of each component.

To give a general estimate of the design quality, for each type of combined spectra, we randomly input 200 spectra from test dataset, which consists of 6,000 samples (20 % of the training dataset). Some representative design cases are shown in Fig. 4. In Fig. 4a, the real images and generated images are presented, and the corresponding simulated spectra (up scattering) and D/U ratios are plotted using blue curves for target spectra and black curves for generated spectra. For up scattering, the presented normalized spectra are what we used into training. While for D/U ratios, the normalized values are used for training but the original logarithmic values are presented in Fig. 4 to convey the physical meaning clearly. Obviously, both the generated scattering spectra and direction ratios accord with the targeted results well. The inverse design based on scattering labels is pretty challenge due to the complex lineshapes made from multipole coupling. Moreover, the direction ratios inherently relate to the phase information of electromagnetic modes, which lead to more complex lineshapes. Therefore, current results can demonstrate DDGAN's capability on inverse design. In Fig. 4b, spectral labels with more contents were used, which combine 160-point up scattering in the visible range (440-787 nm), 80-point D/U ratio (522-600 nm), 80-point F/B ratio (600-707 nm), and 80-point L/R ratio (707-857 nm). The similarity of up scattering and D/U ratio between targeted and generated structures is as good as that in Fig. 4a. While for F/B and L/R ratios, the accuracy drops but still acceptable. Since the lateral scattering (F, B, L, and R) is very sensitive to the structural symmetry as explained in Fig. 2, slight parameter differences would result in drastic changes of lineshapes. Considering the training data volume (30,000) only covers 0.014 % of the amount of ergodic data parameters, further improvement of validation accuracy will be expectable when expanding the dataset. To study the lateral scattering in detail, we further used 320-point L/R ratios in the wavelength range from 440 to 787 nm as spectral labels. The targeted and generated results are shown in Fig. 4c. Although the generated structures are obviously different from the real structures, their L/R ratio curves match well indicating the DDGAN can learn the internal mechanism rather than simply structural imitation. In Fig. 4d, the validation losses (MAE) of the three types of trained networks are presented. The average $MAE_{UP}$ of 200 data is less than 0.15, and the average MAE of direction ratios ($MAE_{D/U}$, $MAE_{L/R}$, $MAE_{F/B}$) can be controlled around 0.05. The reason why $MAE_{L/R}$ of Network 3 (Fig. 4c) is larger than that of Network 1 is because different criterion on data selection. For Network 1, we selected training data and test data with one or more ratios (D/U, L/R, and F/B) with maxima larger than one third of overall maxima or minima less than one third of overall minima. For Network 3, we selected fluctuant data where L/R ratios with maxima larger than one third of overall maxima or minima less than one third of overall minima.

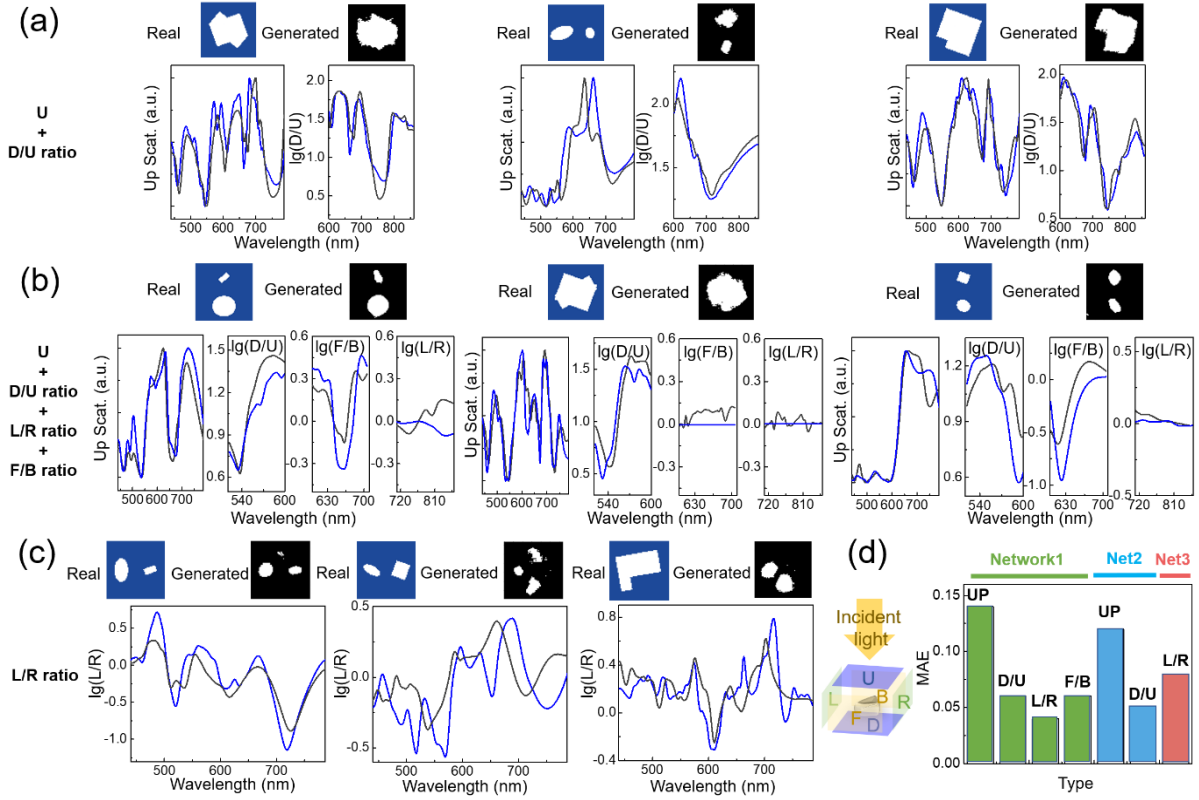

**Figure 4. Evaluation of the generative model using test data.** (a) The training performance using up scattering and D/U ratio: spectral curves and picture background in blue represent the ground-truth, while black represents the generated structures and corresponding spectra. (b) The training performance using up scattering, D/U ratio, L/R ratio, and F/B ratio. (c) The training performance using L/R ratio. (d) The validation losses (MAE) of the three types of trained networks after considering 200 data from test dataset.

Next step, the DDGAN was further utilized to design RGB color routers, and the radiation at full space should be considered rather than specific directions in Fig. 4. Therefore, additional calculations are performed on training dataset to obtain radiation patterns at XZ-plane and then collect radiation angles $\varphi_{max}$ where the strongest radiation happens. Due to the time-consuming far-field calculation, only 20 wavelength points (400-700 nm) were chosen forming as $1\times20$ vector ($\varphi_1$ to $\varphi_{20}$) as shown in Fig. 5a. The 0~360° angle information was further convert to binary $1\times180$ vector, because labels with longer length and fewer discrete values are favorable for training. Using the trained DDGAN, we obtained generated structures with very close wavelength-dependent $\varphi_{max}$ curves compared with targeted one (Fig. 5b). The comparison between real and generated structures shows obvious differences indicating the distinct mechanisms on multipole coupling. In Fig. 5b, we also presented the radiation patterns obtained from real and generated structures at switching wavelengths. The values of $\varphi_{max}$ are consistent but the radiation patterns are different between real structures and generated structures, indicating the same objective was achieved using two different schemes. Since we only fed series of $\varphi_{max}$ into the network and care more about the wavelength-dependent routing behavior, it is common that radiation patterns between real and generated cases are different. Besides the test dataset, we also put some artificial data into the trained network to obtain more functionalized color routers. The principle of dataset generation is shown in Fig. S10a, and in general, we adjusted $\varphi_{max}$ at red and green channels randomly from 90º to 270º and fixed rest of them including the blue channel at 180º. Therefore, in Fig. 5c-e, we presented three types of RGB color routers. Through calculating wavelength-dependent $\varphi_{max}$ of generated structures, we can see pretty good similarity compared with the artificial data. The radiation patterns at RGB channels of each sample are presented together showing the stagger $\varphi_{maxR}$, $\varphi_{maxG}$, and $\varphi_{maxB}$. As talked above, the training process has not considered the complete radiation pattern at a specific wavelength, and in the future, the radiation patterns at specific wavelengths can be easily optimized through a new round of training. In addition, through the duplication of designed unit as an array structure based on lattice Kerker[41], the angular width of the main lobe can be narrowed.

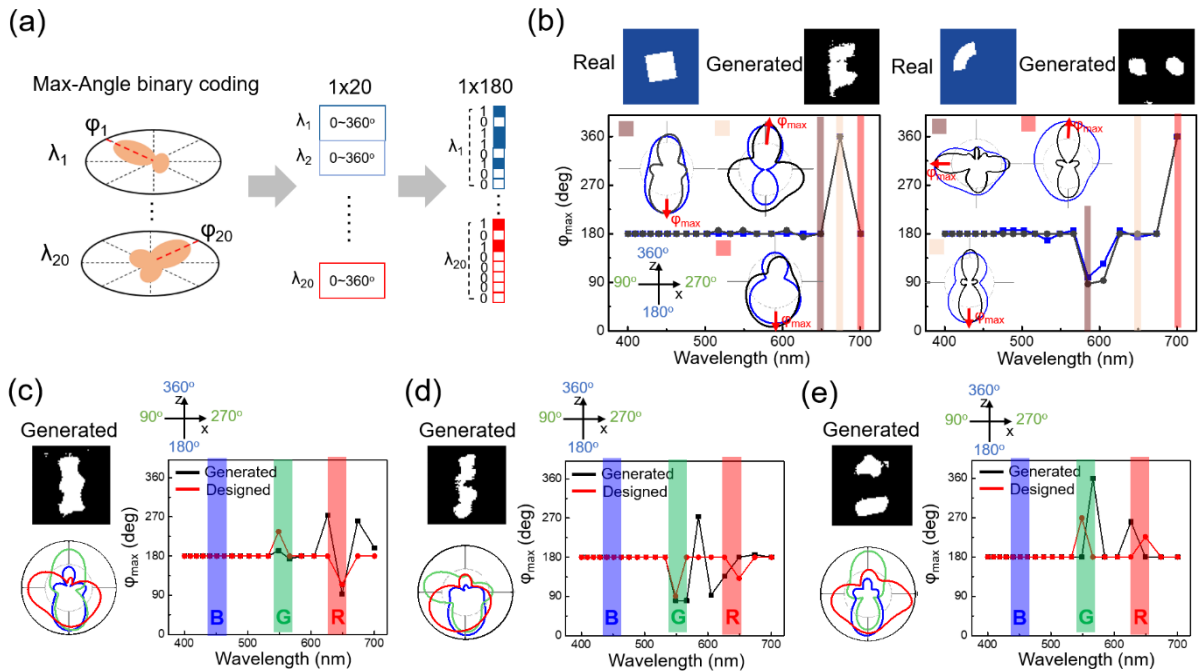

**Figure 5. On-demand inverse design of single-element RGB color routers.** (a) The illustration of max-angle binary coding. (b) Evaluation of the generative model using test data: the ground-truth (generated) geometries, the wavelength-dependent $\varphi_{max}$, and the radiation patterns are in blue (black). $\varphi_{max}$ is labeled by red arrows. (c) The generated structures and corresponding wavelength-dependent $\varphi_{max}$ graphs (black curves) through feeding into designed RGB-stagger $\varphi_{max}$ graphs (red curves). The radiation patterns at RGB channels of each sample are plotted together using red, green, and blue colors.

The unique Kerker switching of dielectric nanostructures makes it possible to design narrowband color routing as illustrated in Fig. 6a. Utilizing the fast switching from left to right, one structure can scatter incident light to different underneath pixels within a narrow wavelength band. To find a structure with the most obvious Kerker switching at desired wavelengths, we utilized the trained DDGAN for 320-point L/R ratio (Fig. 4c) and input artificial Lorentzian lineshapes as shown in Fig. 6b-d. The design principle of Lorentzian lineshapes is explained in Fig. S10b, showing data ranges of topline, center λ, and FWHM. The generated structures as well as corresponding L (R) scattering and L/R ratios are presented in Fig. 6b-d. As we expected, the input Lorentzian lineshapes prompt the generation of Kerker switching channels. The lateral scattering of all three samples shows significant switching performance from left-dominated to right-dominated. Followed the idea in Fig. 1 and 2, we calculated the differential L/R ratio of all three samples as shown in Fig. 6e, and the generated results are comparable to the optimized results from symmetry-breaking nanodisk (Fig. 2c). Those Kerker switching occur at different wavelengths. For comparison, they are plotted together using $\lambda_1$ to $\lambda_n$ to represent the studied regions. More importantly, the generated Kerker switching from DDGAN has larger switching range than the design from parameter sweeping. For No. 3 sample, the L/R ratio (without logarithm) changes from 0.17 at 616 nm to 6.05 at 655 nm. While for the chosen design from parameter sweeping (Channel B in Fig. 2c), the L/R ratio switches from 0.55 at 662 nm to 5.08 at 690 nm. Channel C (λ~800 nm) in Fig. 2 was not used for comparison because of the training process only includes visible range.

We further explore the mechanism behind the best Kerker switching behavior (No. 3). The 3D radiation patterns show a clear switching from right to left (Fig. 6f). Notably, we only presented the top view of 3D radiation patterns to mainly embody the lateral scattering, but the translucent treatment can let us see other parts. Using the spatial multipole decomposition proposed in Fig. 1 and 2, we further calculated radiation patterns contributed by each mode as shown in Fig. S11. However, for this case, we cannot get helpful information to explain the Kerker switching behavior only based on radiation patterns at XY plane (elevation angle $\theta=0°$). Due to the complicated multipole modes with different orientations, radiation patterns considering both elevation and azimuth angles are necessary. Therefore, we improved our code and calculated the 3D radiation pattern of each mode as shown in Fig. 6g. At 616 nm, we can clearly see that EO and EQ mode are dominated, and the orientation of EO along x-axis leads to the enhancement at right direction and the suppression at left direction. While at 655 nm, the EO mode greatly weakened and only EQ is dominated. Moreover, the radiation pattern of EQ mode obviously changes from 616 nm to 655 nm, rendering left-dominated scattering at 655 nm. The simulated near-field distributions in Fig. 6h further confirm the switch from EO to EQ. The EO mode at 616 nm can be regarded as the combination of two resonators' EQs, and the EQ mode at 655 nm only comes from the left Si resonator. To the best of our knowledge, it is the first time to discover the EO-mode related lateral-scattering color routing, demonstrating the core advantage of generative AI on guiding physics research.

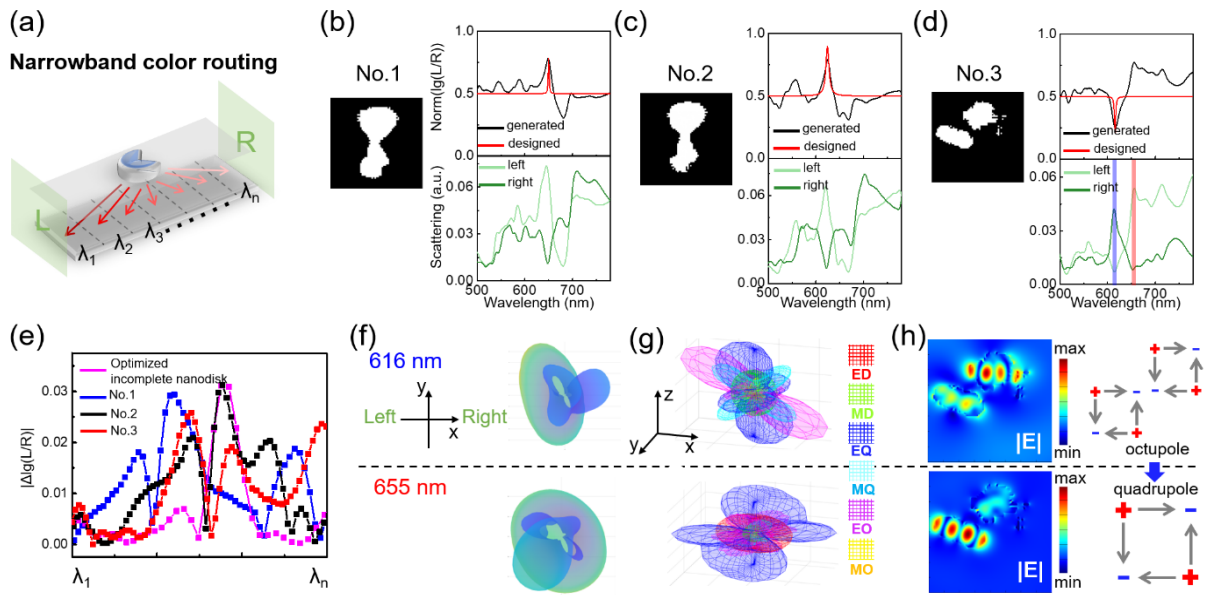

**Figure 6. Design of narrowband color router and the AI-enabled finding of Kerker switching channel.** (a) The concept of narrowband color routing using the Kerker switching from left to right to divide $\lambda_1$ to $\lambda_n$ spatially. (b-d) The generated structures with corresponding L/R ratios and L(R) scattering spectra through feeding into artificial Lorentzian lineshapes located at different wavelengths (red curves).

(e) The differential logarithmic L/R ratios within the switching regions of No. 1-3 samples and the optimized structure in Fig. 2. (f) The top view of 3D radiation patterns switching from right-dominated (616 nm) to left-dominated (655 nm). (g) The mode-resolved 3D radiation patterns at 616 nm and 655 nm. (h) Simulated near-field distributions, and a schematic diagram showing the switching from EO mode (616 nm) to EQ mode (655 nm).

## 3. CONCLUSIONS

In conclusion, leveraging the multipole coupling of dielectric nanostructures, we introduce the concept of Kerker switching for both axial and lateral scattering. This phenomenon enables the 180-degree steering of scattering direction within an extremely narrow wavelength range. However, conventional multipole decomposition methods are inadequate for handling complex scenarios involving multiple directions and wavelengths. To address this limitation, we propose a generative AI model, DDGAN, for the design of multifunctional color routers. This model generates Si structures with desired "combined spectra" composed of scattering spectra from specific directions and direction ratios of multiple directions. We further calculated full-space radiation patterns and fed wavelength-dependent radiation angles into DDGAN. Using this trained network, we successfully design Si structures exhibiting targeted radiation behaviors and RGB color routing capabilities. Lastly, we propose narrowband color routing based on the Kerker switching of lateral scattering. Guided by the trained DDGAN, we identify the Kerker switching channel with the highest switching rate and largest switching range, revealing the dominance of EO mode switching from left to right, a phenomenon that has received little attention before.

## 4. Experimental Section

**Spectral and spatial (2D & 3D) multipole decomposition**

The basic idea is based on Alaee's exact expression[39] and referred MENP matlab code proposed by T. Hinamoto and M. Fujii.[42] First, FDTD simulation is carried out to get wavelength-dependent electric fields (E(x,y,z,f)). So the induced current density (**J**) inside a resonator can be obtained from E(x,y,z,f):

$$\mathbf{J}(\mathbf{r}) = -i\omega\varepsilon_0\left(n^2-1\right)\mathbf{E}(\mathbf{r}) \tag{2}$$

where $\omega$ is the angular frequency, $\varepsilon_0$ is the vacuum permittivity, and $n$ is the refractive index of the dielectric resonator.

Second, the multipole moments including ED ($\mathbf{p_0}+\mathbf{T_2}$), MD ($\mathbf{m_1}$), EQ ($\hat{Q}_1+\hat{Q}_3$), MQ ($\hat{M}_2$), EO ($\hat{O}_2^{(e)}$), MO ($\hat{O}_3^{(m)}$) can be derived from $\mathbf{J}(\mathbf{r})$ as:

$$\mathbf{p_0} = \frac{i}{\omega}\int_{Vs} j_0(k_d r')\mathbf{j}(\mathbf{r}')d\mathbf{r}' \tag{3}$$

$$\mathbf{T_2} = \frac{ik_d^2}{2\omega}\int_{Vs} \frac{j_2(k_d r')}{(k_d r')^2}\left[3(\mathbf{r}'\cdot\mathbf{j})\mathbf{r}' - r'^2\mathbf{j}\right]d\mathbf{r}' \tag{4}$$

$$\mathbf{m_1} = \frac{3}{2}\int_{Vs} \frac{j_1(k_d r')}{k_d r'}[\mathbf{r}'\times\mathbf{j}]d\mathbf{r}' \tag{5}$$

$$\hat{Q}_1 = \frac{3i}{\omega}\int_{Vs} \frac{j_1(k_d r')}{k_d r'}\left[3(\mathbf{r}'\mathbf{j}+\mathbf{j}\mathbf{r}')-2(\mathbf{r}'\cdot\mathbf{j})\hat{U}\right]d\mathbf{r}' \tag{6}$$

$$\hat{Q}_3 = \frac{6ik_d^2}{\omega}\int_{Vs} \frac{j_3(k_d r')}{(k_d r')^3}\left[5(\mathbf{j}\cdot\mathbf{r}')\mathbf{r}'\mathbf{r}' - r'^2(\mathbf{r}'\mathbf{j}+\mathbf{j}\mathbf{r}')-(\mathbf{j}\cdot\mathbf{r}')r'^2\hat{U}\right]d\mathbf{r}' \tag{7}$$

$$\hat{M}_2 = 15\int_{Vs} \frac{j_2(k_d r')}{(k_d r')^2}\left([\mathbf{r}'\times\mathbf{j}]\mathbf{r}' + \mathbf{r}'[\mathbf{r}'\times\mathbf{j}]\right)d\mathbf{r}' \tag{8}$$

$$\hat{O}_2^{(e)} = \frac{15i}{\omega}\int_{Vs} \frac{j_2(k_d r')}{(k_d r')^2}\left(\mathbf{j}\mathbf{r}'\mathbf{r}' + \mathbf{r}'\mathbf{j}\mathbf{r}' + \mathbf{r}'\mathbf{r}'\mathbf{j} - \hat{A}\right)d\mathbf{r}' \tag{9}$$

$$\hat{O}_3^{(m)} = \frac{105}{4}\int_{Vs} \frac{j_3(k_d r')}{(k_d r')^3}\left([\mathbf{r}'\times\mathbf{j}]\mathbf{r}'\mathbf{r}' + \mathbf{r}'[\mathbf{r}'\times\mathbf{j}]\mathbf{r}' + \mathbf{r}'\mathbf{r}'[\mathbf{r}'\times\mathbf{j}] - \hat{A}'\right)d\mathbf{r}' \tag{10}$$

where $k_d$ is the wavenumber, $\mathbf{r}'$ is the position vector in spherical integral, $j_n(\rho)$ is the spherical Bessel function of n order.

Using the multipole expressions above, the total scattering cross section can be calculated as:

$$\begin{aligned}P_{sca} \simeq &\frac{k_0^4}{12\pi\varepsilon_0^2\upsilon_d\mu_0}|\mathbf{p_0}+\mathbf{T_2}|^2 + \frac{k_0^4\varepsilon_d}{12\pi\varepsilon_0\upsilon_d}|\mathbf{m_1}|^2 + \frac{k_0^6\varepsilon_d}{1440\pi\varepsilon_0^2\upsilon_d\mu_0}\sum_{\alpha\beta}|Q_{1\alpha\beta}+Q_{3\alpha\beta}|^2\\ &+\frac{k_0^6\varepsilon_d^2}{160\pi\varepsilon_0\upsilon_d}\sum_{\alpha\beta}|M_{2\alpha\beta}|^2 + \frac{k_0^8\varepsilon_d^2}{3780\pi\varepsilon_0^2\upsilon_d\mu_0}\sum_{\alpha\beta\gamma}|O_{2\alpha\beta\gamma}^{(e)}|^2 + \frac{k_0^8\varepsilon_d^3}{3780\pi\varepsilon_0\upsilon_d}\sum_{\alpha\beta\gamma}|O_{3\alpha\beta\gamma}^{(m)}|^2\end{aligned} \tag{11}$$

However, the expression for scattering cross section only solves the spectral decomposition. For spatial decomposition, we should go back to the scattered electric field contributed by each mode:

$$\mathbf{E_n}(\mathbf{r}) \simeq k_0^2 \frac{e^{ik_d r}}{\varepsilon_0 4\pi r} \begin{pmatrix} \left[\mathbf{n}\times\left[(\mathbf{p_0}+\mathbf{T_2})\times\mathbf{n}\right]\right] + \frac{1}{\upsilon_d}\left[\mathbf{m_1}\times\mathbf{n}\right] + \frac{ik_d}{6}\left[\mathbf{n}\times\left[\mathbf{n}\times\left(\hat{Q}_1+\hat{Q}_3\right)\mathbf{n}\right]\right] \\ + \frac{ik_d}{2\upsilon_d}\left[\mathbf{n}\times\hat{M}_2\mathbf{n}\right] + \frac{k_d^2}{6}\left[\mathbf{n}\times\left[\mathbf{n}\times\hat{O}_2^{(e)}\mathbf{nn}\right]\right] + \frac{k_d^2}{6\upsilon_d}\left[\mathbf{n}\times\hat{O}_3^{(m)}\mathbf{nn}\right] \end{pmatrix} \quad (12)$$

where $\mathbf{n}=\mathbf{r}/r$ is the unit vector, $k_0$ is the wavenumbers in vacuum. Through converting above expression into spherical coordinates, we can obtain the far-field at arbitrary 2D planes or 3D space contributed by each mode.

**Inverse design using DDGAN**

The DDGAN is based on the typical generative model (WGAN-GP) with conditional labels. Using the PyTorch framework, the DDGAN is constructed by a generator (five transposed convolutional layers) and a discriminator (five convolutional layers). The detailed network is shown in Fig. S9. The training dataset consists of 30,000 Si structures within 800x800 nm$^2$ region. Structural designs contain ellipses, rectangles, L-shapes, arcs, crosses, parallel dimers, and vertical dimers to sufficiently describe the scattering behaviors of Si nanostructures. The size parameters are looped over and randomly picked. The height of Si structures is fixed at 200 nm, and no substrate is set to simplify analysis. The sectional views are represented by 64x64 pixel images.

FDTD simulations were performed on each structure using total-field scattered-field (TFSF) incident light with polarization along x-axis (-x for left & +x for right), and 600-point scattering spectra of all six planes in the wavelength range from 400-1600 nm were collected. Further slicing and combination processes produce combined spectra for different training targets. Different parts of combined spectra are normalized separately, and scattering spectra and ratios use different normalized method: scattering spectra were normalized one by one, while direction ratios were normalized globally using the maximum and minimum of whole training dataset. For combined spectra with direction ratios, further data selection was performed to exclude ineffective design. For example, for the training shown in Fig. 4b, we excluded the training data if no ratios (D/U, L/R, and F/B) are significant (significant means that maxima are larger than one third of overall maxima or minima are less than one third of overall minima). During the training process, the objective function:

$$L = \mathop{\mathrm{E}}_{\tilde{x} \sim P_g}\left[D(\tilde{x})\right] - \mathop{\mathrm{E}}_{x \sim P_r}\left[D(x)\right] + \lambda \mathop{\mathrm{E}}_{\hat{x} \sim P_{\hat{x}}}\left[\left(\left\|\nabla_{\hat{x}} D(\hat{x})\right\|_2 - 1\right)^2\right] + \mu \sum_n \left|S - D^{'}(\tilde{x})\right|/n \qquad (13)$$

is optimized using Adam optimizer, where $P_{\hat{x}}$ is the sampling uniformly along straight lines between pairs of points sampled from the data distribution $P_r$ and the generator distribution $P_g$. $D^{'}$ mean the other form of the Discriminator's output with the same dimension of labels (S). $\lambda = 10$ is the gradient penalty factor. $\mu = 100$ is the label loss factor. The validation losses (MAE), defined as $\mathrm{MAE} = \sum_{N=1}^{N_{\max}} \sum_{\lambda=1}^{\lambda_{\max}} \left(I_{N,\lambda} - \hat{I}_{N,\lambda}\right) / N_{\max} \lambda_{\max}$, are used to compare the performances of different training models. $I_{N,\lambda}$ is the scattering intensity or direction ratio of generated structures, while $\hat{I}_{N,\lambda}$ is the targeted scattering intensity or direction ratio. N means the number of test data, and λ means the number of wavelength points. Notably, the generated combined spectra are divided as how they combined, and MAEs are calculated separately.

**Supporting Information**

Explanations of Kerker switching regarding the evolution of axial scattering (Up and Down), explanations of Kerker switching regarding the evolution of lateral scattering (Left and Right), detailed information of the training dataset, detailed information of the DDGAN architecture, comparisons among DDGAN, DCGAN, and WGAN-GP, design of artificial input labels for RGB and narrowband color routing, the limitation of 2D radial multipole decomposition.

**Conflict of Interest**

The authors declare no conflict of interest.

**ACKNOWLEDGMENTS**

The work was supported by the National Natural Science Foundation of China (No. 62005096) and the Guangdong Basic and Applied Basic Research Foundation (No. 2023B1515020046).

# Supporting Information

# Design of multifunctional color routers with Kerker switching using generative adversarial networks

Jiahao Yan[1]\*, Dayu Zhu[2], Yanjun Bao[1], Qin Chen[1], Baojun Li[1], & Wenshan Cai[2]\*

*1.Guangdong Provincial Key Laboratory of Nanophotonic Manipulation, Institute of Nanophotonics, Jinan University, Guangzhou 511443, China*

*2.School of Electrical and Computer Engineering, Georgia Institute of Technology, Atlanta, Georgia 30332, USA*

\*Corresponding author: jhyan@jnu.edu.cn & wcai@gatech.edu

**Contents**:

## 1. Explanations of Kerker switching regarding the evolution of axial scattering (Up and Down)

In Fig. S1, more spectral information of the silicon nanodisks is presented. As we mentioned in the main article, the height is fixed at 200 nm, and diameters vary from 100 nm to 600 nm. The visualization of up and down scattering as diameter and wavelength change shown in Fig. S1a and b clearly indicates the linearly red-shift of each mode. The distinct differences between up and down scattering confirm the existence of Kerker scattering. The dominated electromagnetic modes are labelled in Fig. S1b, which refers the multipole decomposition in the main article. We also calculated the differential up and down scattering as shown in Fig. S1c and d, visualizing the changing rate driven by Kerker switching. Considering both the differential scattering and the differential direction ratio ($\Delta \lg(Scat_D/Scat_U)$), an ideal Kerker switching channel can be found.

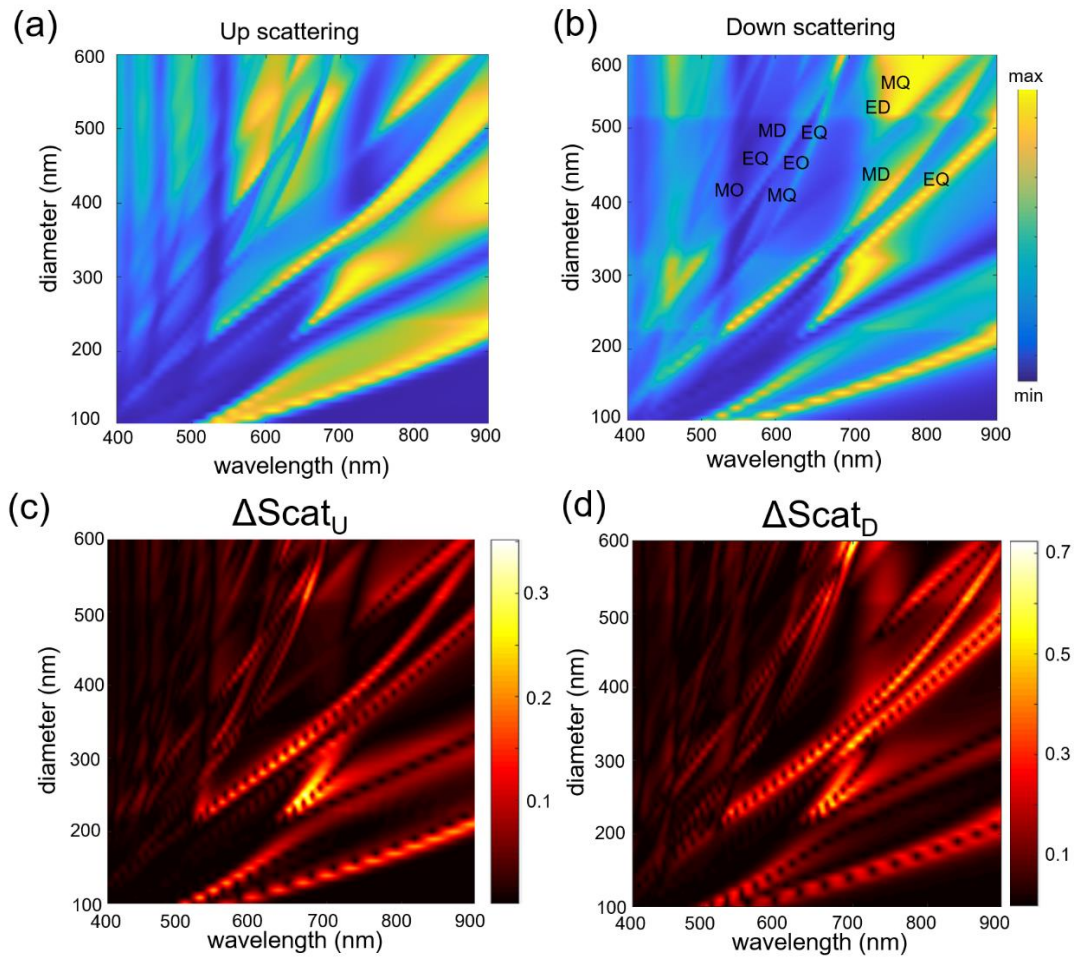

**Figure S1.** The study of axial scattering (up and down) via parameter-sweeping of Si nanodisks. (a, b) Visualization of up (a) and down (b) scattering varied with diameters and wavelengths. (c, d) Visualization

of differential up (c) and down (d) scattering varied with diameters and wavelengths.

As mentioned in the main article, we chose the Si nanodisk with a diameter of 430 nm to study A, B, and C switching channels. Besides Fig. 1d, the 3D radiation patterns of A and C channels are presented in Fig. S2. The values of x, y, and z axes are set consistent, and we can clearly see the dramatic increase and then decrease of down scattering for both A and C channels. Based on the spectral and spatial multipole decomposition in Fig. 1c and e, we can conclude that the switching channel A arises from the sudden appearance of MQ and EO modes at 626 nm, and the channel C arises from the sudden appearance of EQ mode at 827 nm.

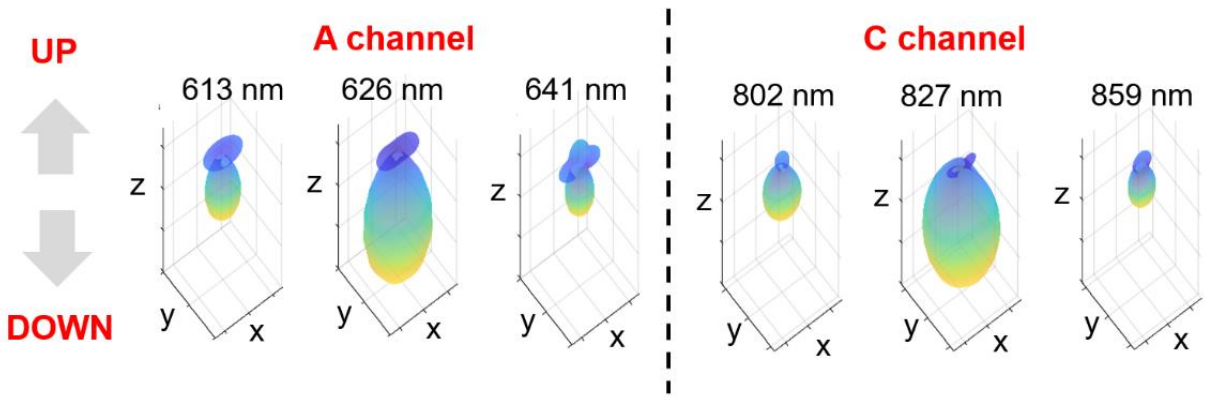

**Figure S2.** The 3D radiation patterns of A channel switching from 613, 626 to 641 nm and C channel switching from 802, 827 to 859 nm. The specific locations of A and C channels are shown in Fig. 1b.

Besides the mode-resolved radiation patterns in XZ plane shown in Fig. 1e, the mode switching in these three channels is also witnessed through multipole decomposition in YZ plane (Fig. S3). The combination of radiation patterns in both XZ and YZ planes gives a more complete picture of each mode's dipole moment and orientation, Since the symmetric structures generate multipole moments simply along x, y, or z directions, the whole picture of multipole modes can be easily expected through prior experiences.[1]

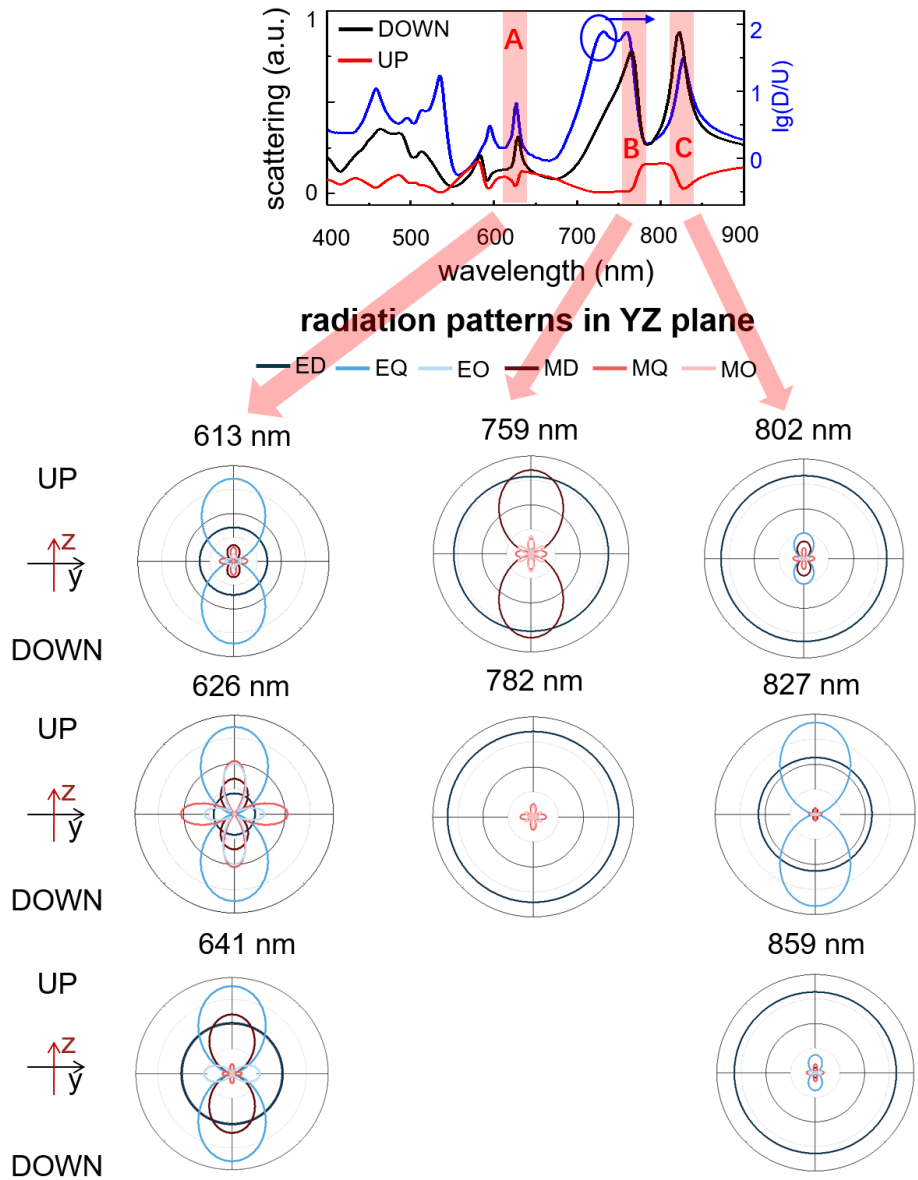

**Figure S3.** Up and down scattering and the corresponding D/U ratio of the chosen Si nanodisk ($d$=430 nm). Radiation patterns in YZ-plane contributed by each mode at three channels are also presented.

## 2. Explanations of Kerker switching regarding the evolution of lateral scattering (Left and Right)

In this section, we provide more lateral scattering characterizations as a supplement of Fig. 2a in the main article. The visualization of scattering intensities (Left and Right) is presented in Fig. S4a and b changing with wavelengths and the notch angles. Notably, the scattering intensities were globally normalized, and we can clearly see the enhancement of scattering intensity when more degree of symmetry-breaking is generated. In general, multipole modes experience blue-shift, but some peak lines are terminated when the notch angle reaches some degree, indicating the difficulties on tracking multipole modes of lateral scattering. The logarithmic direction ratios are presented in Fig. S4c, where we can observe crossovers of some peak lines. That means the variations of modes no matter with notch angle or with wavelength are out-of-step.

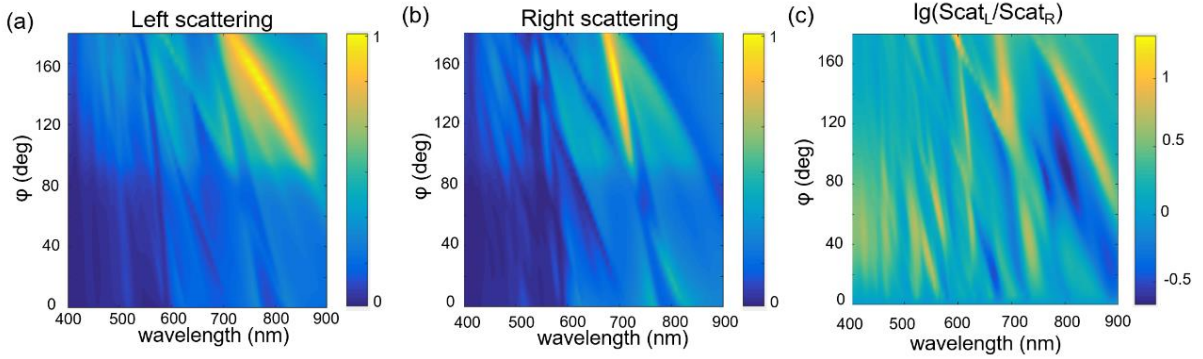

**Figure S4.** (a, b) The left (a) and right (b) scattering intensities changing with wavelengths and notch angles. (c) The logarithmic direction ratios changing with wavelengths and notch angles. The diameter and height of the incomplete Si nanodisk are fixed at 430 nm and 200 nm, respectively.

Another question of symmetry-breaking nanodisks is the simultaneous degeneration of D/U directivities as shown in Fig. S5a and b. When increasing the notch angle, the resonant modes that determine the Kerker scattering broaden and shift away, so some peak lines disappear especially at λ>600 nm. The degenerated modes fail to obtain enough switching rate ($\Delta$lg($Scat_D$/$Scat_U$)) as shown in Fig. S4b. Therefore, the simultaneous design of both D/U and L/R ratios is a challenge.

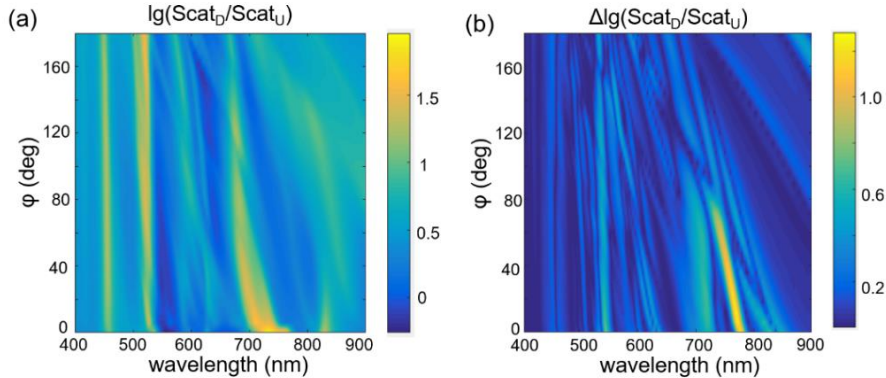

**Figure S5.** (a) The logarithmic D/U direction ratios changing with wavelengths and notch angles. (b) The differential logarithmic D/U direction ratios changing with wavelengths and notch angles. The diameter and height of the Si nanodisk are fixed at 430 nm and 200 nm, respectively.

Besides the radiation patterns and vector diagrams of Channel C in the main article, we also plotted those of Channel A and B in Fig. S6. Interestingly, the Kerker switching of both channels comes from the switching of MQ mode. For A channel, the sudden enhancement of MQ mode at 615 nm explains the left-dominated scattering. From the vector diagrams at 615 nm, we can clearly see the counterbalance between MQ and ED&EQ on the right while the unbalanced coupling between ED and MQ&EQ on the left. For the vector diagrams at 634 nm, the ED mode is dominated on both left and right direction regardless of the 180º phase shift. Therefore, there is no obvious difference between left and right scattering. Similarly, for Channel B, we can explain the right-dominated scattering at 662 nm. The key factor is the 180º phase shift of MQ mode from left to right and the steady EQ mode with no phase change.

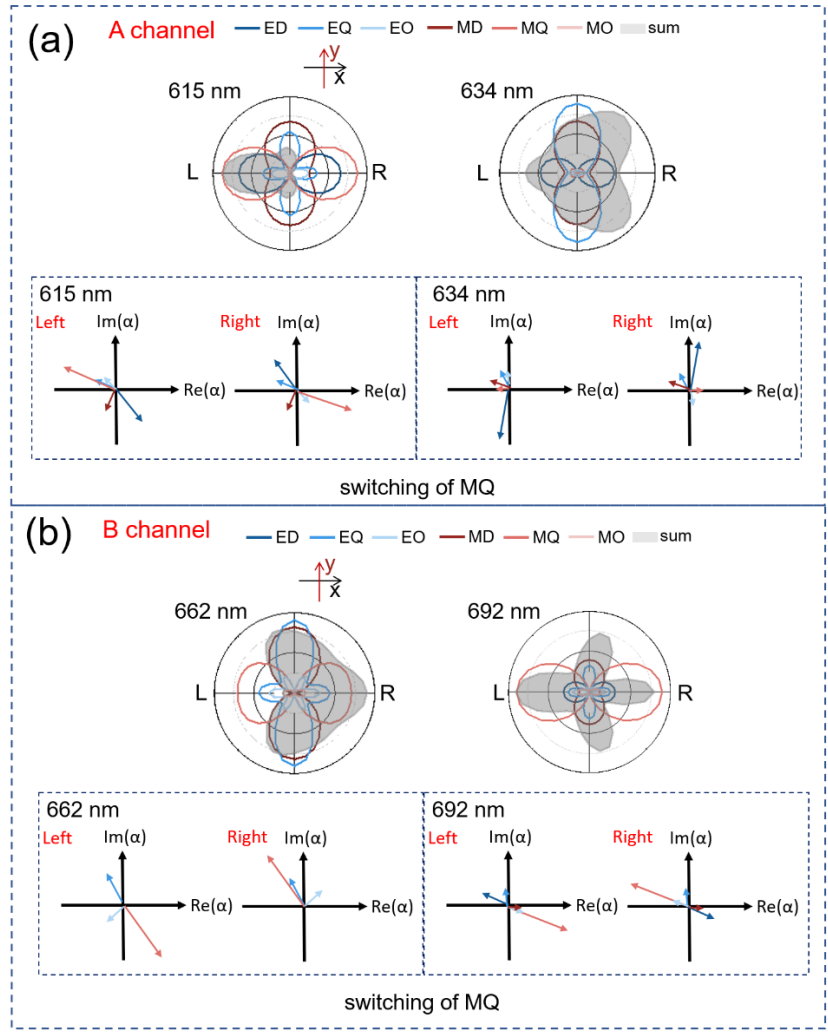

**Figure S6.** (a, b) XY-plane mode-resolved radiation patterns of A (a) and B (b) channels, and vector diagrams of left and right directions at each wavelength.

Different from previous works[2-3] that used multipole moments of specific axis under Cartesian coordinates to calculate the far-field radiation, in this work, we provided a more generalized way to calculate all components of each mode under spherical coordinates. In Fig. S7, we provided the real part of $E_r$, $E_\varphi$, and $E_\theta$ with varied azimuth angle $\varphi$. Obviously, for left and right direction, we only need to consider $E_\varphi$ for calculating amplitude and phase.

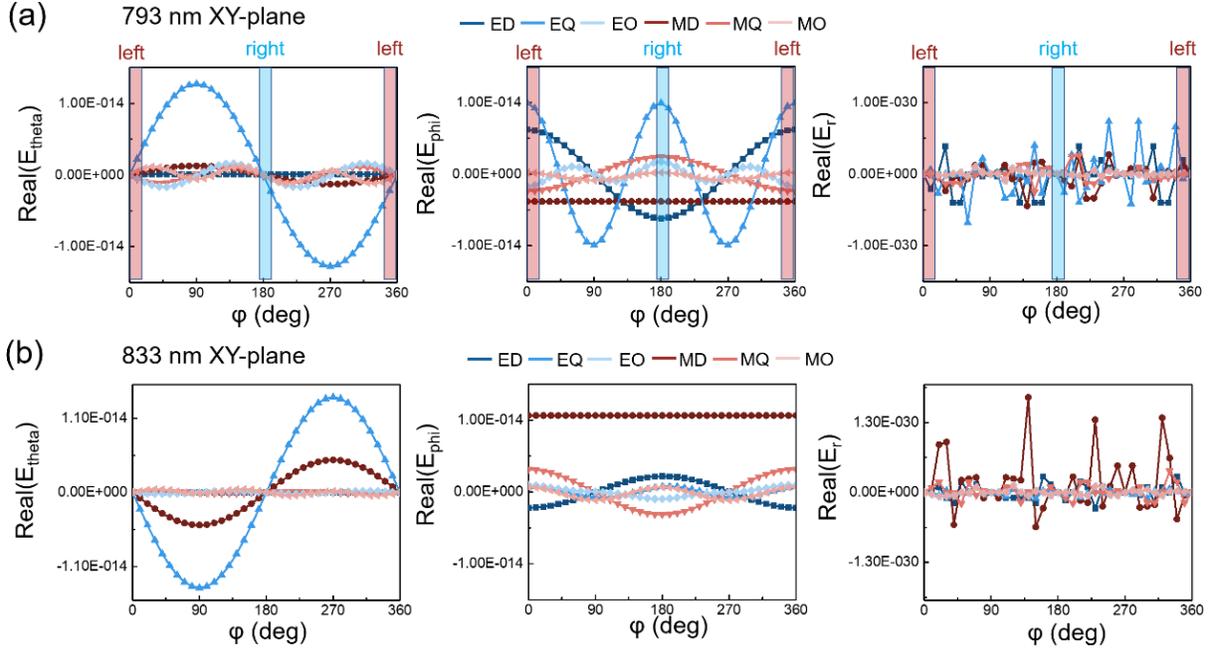

**Figure S7.** (a, b) The real part of $E_r$, $E_\varphi$, and $E_\theta$ with varied azimuth angle $\varphi$ and fixed elevation angle $\theta=0°$ at 793 nm (a) and 833 nm (b).

## 3. Detailed information of the training dataset

A dataset with different classes of geometries was constructed as explained in Fig. S8. All these seven types are widely used in previous studies of all-dielectric nanostructures, and the parameter sweeping is large enough to consider diverse resonant modes and situations of symmetry breaking. In Fig. S8, we use the format as (start, end, step length) to describe each parameter's variation range. Going through each parameter, the number of parameters for ellipses, rectangles, L-shapes, arcs, crosses, parallel dimers, and vertical dimers is 25281, 25281, 10074276, 246240, 10969344, 14112, and 14112, respectively. And those parameters are randomly picked for simulation. The number of used parameters is 4000, 4000, 6000, 6000, 6000, 2000, and 2000 in the order above. For some geometries, the selected parameters only cover a small part of the whole dataset, therefore, the design performance for structures which need careful control on symmetry could be improved if we expand the dataset. Before feeding the parameters into simulation, we also removed few shapes that expand beyond the 800x800 nm simulation region.

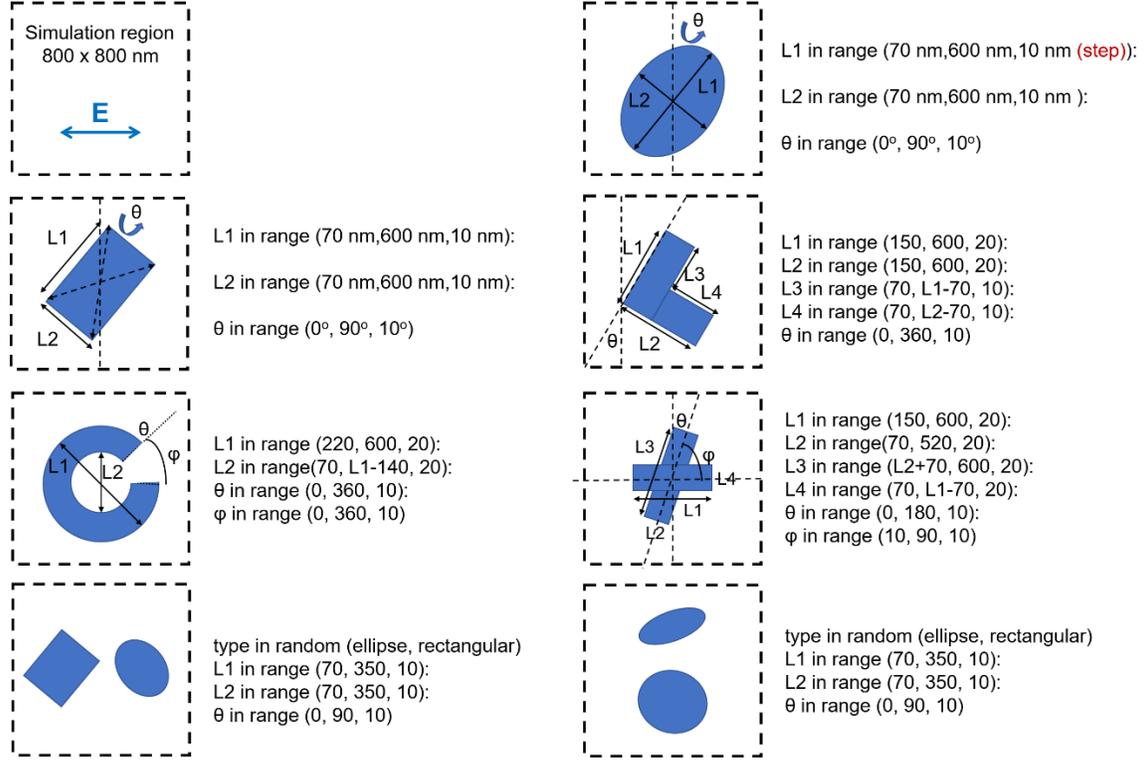

**Figure S8.** Dimension parameters of training dataset.

## 4. Detailed information of the DDGAN architecture

The DDGAN is formed by a generator and a discriminator as shown in Fig. S9. The generator contains five transposed convolutional layers (with N (length of combined spectra), 1024, 512, 256, and 128 input channels or feature maps), and the discriminator is formed by five convolutional layers (with 2, 64, 128, 256, and 512 feature maps). Kernel size is 6x6 in both transposed convolutional and convolutional layers. Each transposed convolutional layer in the generator is followed by a batch normalization and ReLU activation layer except the final layer, where a Tanh activation is used. In the discriminator, each convolutional layer is followed by an instance normalization and leaky ReLU. For the final double output, one is a scalar for Wasserstein distance, and the other is 1xN output with a Sigmoid activation and a MAE calculation. The input of generator is a product of N-point combined spectra and N-point latent vectors, while the input of discriminator is a combination of two 64x64 maps: one is the real or generated images, and the other is reshaped combined spectra after passing through a fully-connected layer. The combined spectra also fed into the end of discriminator to calculate the MAEs. The training process took roughly 1h 30min using an NVIDIA GeForce RTX3080 GPU.

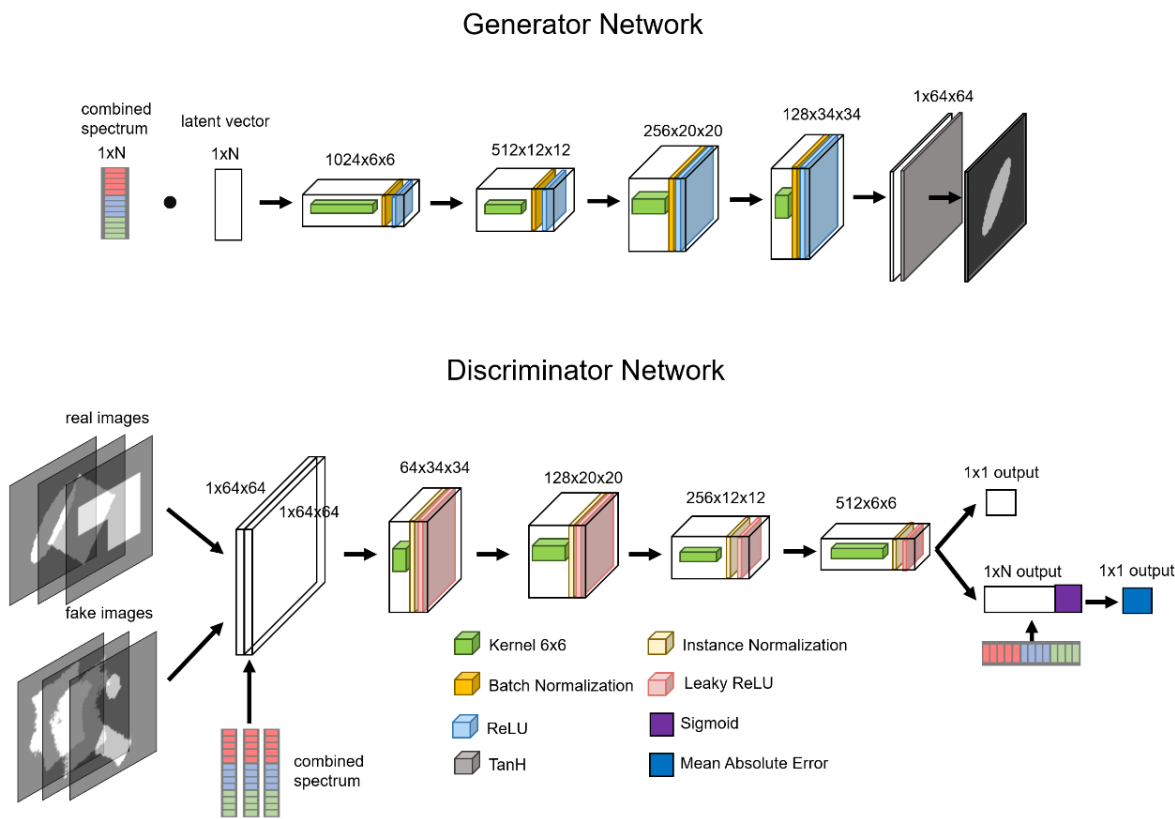

**Figure S9.** The architectures of generator and discriminator in the proposed DDGAN model.

## 5. Comparisons among DDGAN, DCGAN, and WGAN-GP

In the main article, we used several types of spectral labels. In this part, we fix the spectral label and evaluate the training effects of DDGAN, DCGAN and WGAN-GP. The input label is the combination of 160-point Up scattering in the visible range and 160-point D/U ratio. The hyperparameter tuning was conducted for each model, and the key parameters are presented in Table S1. Compared with DCGAN, WGAN-GP add gradient penalty for optimizing to avoid gradient explode or vanish. Furthermore, based on WGAN-GP, DDGAN add label loss optimization with a factor of 100. Consequently, the validation loss (MAE) for up scattering is 0.195, 0.171, and 0.118 for DCGAN, WGAN-GP, and DDGAN, respectively. In addition, the MAE for D/U ratio is 0.105, 0.082, and 0.052 in the order above.

| Model | Batch size | Learning rate | Optimizer | Epoch | Generator | Discriminator | Gradient penalty factor | Label loss factor | Validation loss (MAE) for U scattering | Validation loss (MAE) for D/U ratio |
|---|---|---|---|---|---|---|---|---|---|---|
| DCGAN | 64 | 0.0001 | Adam β1=0 β2=0.999 | 200 | ConvTranspose+BatchNorm+ReLU+…+Tanh | Conv+LeakyReLU+BatchNorm+…+Sigmoid | NA | NA | 0.195 | 0.105 |
| WGAN-GP | 64 | 0.0001 | Adam β1=0 β2=0.999 | 200 | ConvTranspose+BatchNorm+ReLU+…+Tanh | Conv+InstanceNorm+LeakyReLU+… | 10 | NA | 0.171 | 0.082 |
| DDGAN | 64 | 0.0001 | Adam β1=0 β2=0.999 | 200 | ConvTranspose+BatchNorm+ReLU+…+Tanh | Conv+InstanceNorm+LeakyReLU+… (+Sigmoid) | 10 | 100 | 0.118 | 0.052 |

Label: combined spectra (160-point U scattering+ 160-point D/U ratio)

**Table S1.** The optimized hyperparameters and the validation losses of three models.

## 6. Design of artificial input labels for RGB and narrowband color routing

As elaborated in the main article, the simultaneous collection of scattering spectra from six planes can provide us the wavelength-dependent radiation patterns. To reduce complexity, we only extracted the angle where maximum radiation occurs ($\varphi_{max}$) at each wavelength point. In Fig. S10a, we give schematic diagram showing the polarization direction, the definition of left and right, and the position of angular coordinate. Since $\varphi_{max}=180°$ (down-dominated) is the most common, we fixed $\varphi_{max}$ of most wavelength points at 180° and only adjusted $\varphi_{max}$ at green and red bands. In practical application, different $\varphi_{max}$ at RGB three channels directs the constructing of RGB color routers. To reduce the overlap between each channel, the random generated wavelength-dependent $\varphi_{max}$ should follow these restrictions: $\varphi_B=180°$, $|\varphi_G-180°|>30°$, $|\varphi_R-180°|>30°$, and $|\varphi_R-\varphi_G|>30°$. The 20-point angle information is further converted to 180-point binary data.

For the design of narrowband color router, we studied the switching from L to R and tried to find the fastest switching through feeding narrowband Lorentzian lineshapes as L/R ratios as shown in Fig. S10b. Based on Fig. 4c in the main article, we are aware that the L/R ratio spectra are fluctuant. Therefore, a single Lorentzian lineshape is not precise enough as a label to prompt the structural generation. To solve this problem, we generated curves with ranges of topline $y_c$, center $\lambda$, and FWHM $w$. In addition, the picked spectra are repeated several times and packaged as a 200xN matrix for prediction. The data selection is also needed after the prediction and evaluation process to choose structures with the best performance.

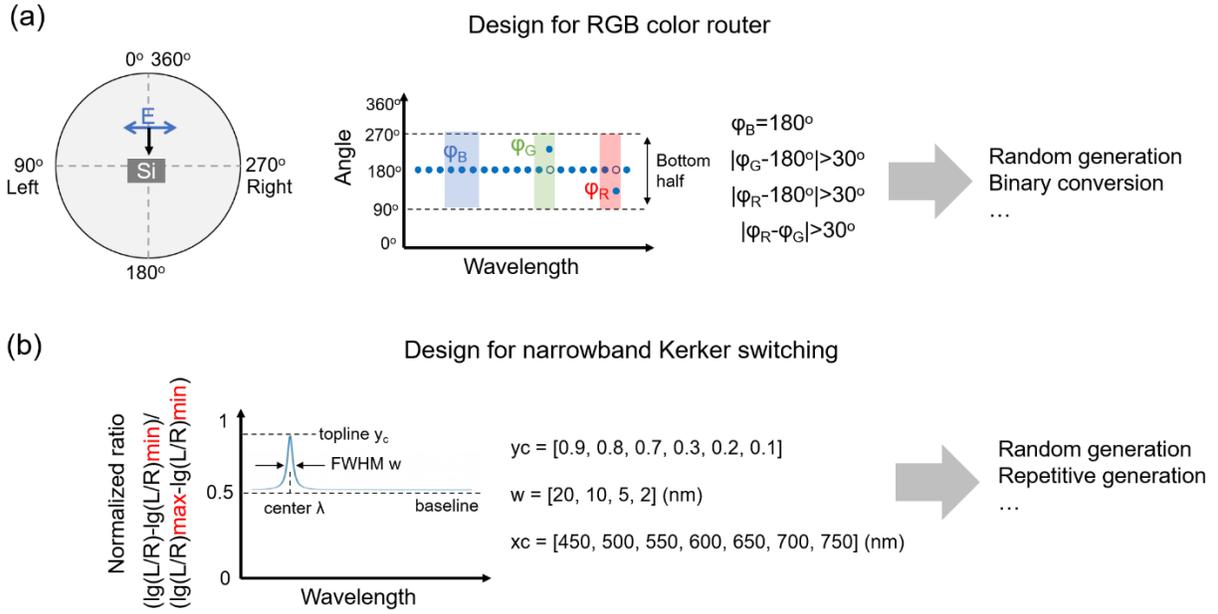

**Figure S10.** (a, b) The workflow showing the input data for the inverse design of RGB color routers (a) and narrowband color routers (b).

## 7. The limitation of 2D radial multipole decomposition

For No.3 structure in Fig. 6, we plot the 2D radiation pattern contributed by each mode as shown in Fig. S11. Compared with the improved 3D radiation pattern in the main article, here we illustrated that 2D mode-resolved radiation patterns are not able to analysis the directional switching of lateral scattering. At 616 nm, the right scattering is supposed to be dominated. However, the in-phase relationship between EQ and MD mode at both left and right directions means no clue for unidirectional scattering. Similarly, at 655 nm, although we can explain the directional effect through phase differences between EO and EQ modes, the orientation of EQ mode mismatches the actual 3D radiation pattern shown in Fig. 6f. Therefore, in this part, we want to manifest that the spatial multipole decomposition of symmetry breaking structures depends on both azimuth angle and elevation angle. Only analyzing XY-plane with fixed elevation angle $\theta=0°$ is not enough.

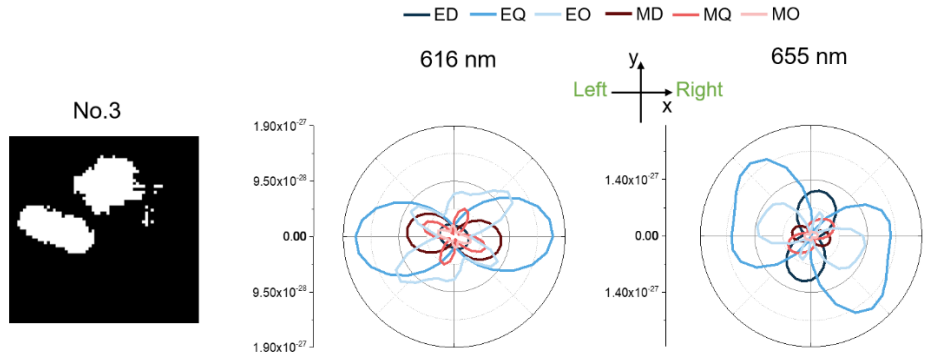

**Figure S11.** The generated structures for narrowband Kerker switching, and the corresponding XY-plane mode-resolved radiation patterns at 616 and 655 nm.